\documentclass[aps,prb,reprint,showkeys]{revtex4-1}

\usepackage{subcaption}
\usepackage{bm,graphicx,tabularx,array,booktabs,dcolumn,xcolor,microtype,multirow,amscd,amsmath,amssymb,amsfonts,physics,siunitx,enumitem,upgreek}
\usepackage[version=4]{mhchem}
\usepackage[normalem]{ulem}
\usepackage{mleftright}
\usepackage{wrapfig}

\usepackage[utf8]{inputenc}
\usepackage[T1]{fontenc}

\AtBeginDocument{\nocite{achemso-control}}

\usepackage{tikz}
\usetikzlibrary{arrows.meta,positioning,shapes.misc,arrows}

\newcommand{\br}{\bm{r}}

\newcommand{\xc}{\text{xc}}

\usepackage{mathtools}

\newcommand{\Vn}{V_\text{nn}}

\newcommand{\cA}{\mathcal{A}}
\newcommand{\cC}{\mathcal{C}}
\newcommand{\cP}{\mathcal{P}}
\newcommand{\cQ}{\mathcal{Q}}

\newcommand{\cR}{\mathcal{R}}

\newcommand{\Ns}{m}

\definecolor{hughgreen}{RGB}{100, 0, 140}

\allowdisplaybreaks

\usepackage[
	colorlinks=true,
    citecolor=blue,
    linkcolor=blue,
    filecolor=blue,      
    urlcolor=blue,	
    breaklinks=true
	]{hyperref}
\newcommand{\UCL}{Department of Chemistry, University College London, London, WC1H 0AJ, U.K.}

\begin{document}	

\title{Density functional theory for general restricted open-shell configurations}
\title{Restricted open-shell Kohn--Sham theory for arbitrary genealogical spin coupling}

\date{\today}
\author{Hugh G.~A.~\surname{Burton}}
\email{h.burton@ucl.ac.uk}
\affiliation{\UCL}
\author{Cameron Y.~Z.~\surname{Brown}}
\affiliation{\UCL}

\begin{abstract}
We introduce a generalization of restricted open-shell Kohn--Sham (ROKS) theory to a
single configuration state function (CSF) with arbitrary genealogical spin coupling.
This CSF-ROKS theory establishes a spin-pure density functional extension of the recently developed CSF-based restricted open-shell
Hartree--Fock (CSF-ROHF) theory, providing a computationally efficient route to incorporate dynamic electron correlation.
The energy is expressed as a weighted sum over unrestricted determinants built from a common set of restricted orbitals,
and the number of determinants grows quadratically with the number of open shells.
Numerical calculations for \ce{N2} and the polyacene series show that CSF-ROKS solutions retain
chemically interpretable  character.
Exchange-correlation functionals preferentially stabilize low-seniority configurations, shifting the onset of
polyradical character in polyacenes to longer chains than predicted by CSF-ROHF.
Furthermore, CSF-ROKS can be used to target state-specific spin-pure excitations in the vinyl
and allyl radicals, and the \ce{[Cr(ddpd)2]^{3+}} molecular ruby.
However, CSF-ROKS incorrectly predicts ferromagnetic coupling in a set of bridged transition metal dimers, 
indicating that kinetic exchange requires more advanced correlated methods.
\end{abstract}

\maketitle


\section{Introduction}
\label{sec:intro}

Modeling open-shell electronic states is essential to understand the chemistry of 
molecular magnets,\cite{Chilton2022} transition metal complexes,\cite{Malrieu2014} 
organic radicals,\cite{Krylov2017} and excited states.\cite{Gonzalez2011} 
However, obtaining accurate and spin-pure predictions of low-energy open-shell states is complicated by
the presence of many near-degenerate electron configurations.
Wavefunction methods rely on complete active space 
self-consistent field (CASSCF) theory\cite{Roos1980a,Roos1980b,RoosBook} 
and its dynamic correlation extensions,\cite{Roos1995,Andersson1990,Andersson1992,Hirao1992,Finley1998,Malrieu1995,Angeli2001,Angeli2002,Lyakh2011,Shiozaki2011,Szalay2012,Manni2014,Evangelista2018}
but these methods scale exponentially with the number of unpaired electrons.
In contrast, Kohn--Sham density functional theory (KS-DFT) 
retains a single-determinant picture with low-order polynomial scaling,\cite{Hohenberg1964,Kohn1965,Becke2014} 
but relies on spin-polarized densities to give accurate open-shell energies.\cite{Pople1995,Cohen2008b,Perdew2009,Jacob2012,Verma2020}
New methods that achieve quantitative accuracy, preserve spin symmetry, 
and retain low polynomial scaling are urgently needed.

Recent work has shown that many low-spin open-shell states are dominated by a 
single configuration state function (CSF) when represented using localized orbitals.\cite{Dobrautz2019,Manni2020,Manni2021a,Manni2021b,Izsak2023,MartiDafcik2024,MartiDafcik2025}
In these cases, the multiconfigurational character arises from a fixed spin-coupling pattern
and the determinant expansion is defined analytically by the genealogical Clebsch--Gordan coefficients,
avoiding the need for multiconfigurational optimization altogether.
For example, the ground state of the stretched \ce{N2} molecule can be approximated by a single CSF 
with two high-spin quartets localized on each \ce{N} atom coupled to an overall singlet, whereas the 
equivalent active-space representation includes 400 determinants.\cite{MartiDafcik2024,Manni2020}
It has been shown that using the corresponding many-body CSF basis can
increase the Hamiltonian sparsity to improve the performance of approximate 
configuration interaction (CI) solvers,\cite{Dobrautz2019,Manni2020,Manni2021a,Dobrautz2022} 
such as full CI quantum Monte Carlo,\cite{Booth2009}
and creates a block-diagonal structure that can be leveraged to target excited states.\cite{Manni2021b,Manni2023,Song2024}
This compression of apparent multiconfigurational character into a single CSF using
localized orbitals presents a conceptual opportunity to develop computationally efficient 
methods for low-spin open-shell states.

Motivated by the accuracy of a single CSF for spin-coupled states, 
a generalized restricted open-shell Hartree--Fock (ROHF) theory has been developed to optimize the CSF orbitals.
In this so-called ``CSF-ROHF'' approach, 
the molecular orbitals are optimized for a fixed spin-coupling pattern that is pre-defined 
using the genealogical coupling scheme.
Several SCF optimization schemes have been proposed based on
Fock diagonalization,\cite{Gouveia2024} geometric direct minimization,\cite{Burton2025} and super-CI.\cite{Song2026}
In particular, geometric direct minimization provides robust energy minimization using a quasi-Newton strategy
and can be implemented with $\mathcal{O}(n^4 m)$ scaling, where $n$ and $m$ are the number of 
atomic orbitals and distinct orbital shells respectively.
CSF-ROHF theory allows low-spin open-shell configurations to be optimized with mean-field computational scaling, 
and has been shown to successfully provide orbital-optimized low-spin configurations in antiferromagnetic 
transition metal clusters, such as the \ce{[Fe2S2(SCH3)_4]^{2-}} complex, and organic polyacene chains.\cite{Gouveia2024,Burton2025,Song2026}

As a mean-field theory, CSF-ROHF  lacks the additional electron correlation required for quantitative accuracy.
For example, it fails to predict the correct ordering of spin states for antiferromagnetic ground states in bridged transition metal dimers.\cite{Gouveia2024,Burton2025,Song2026}
We therefore need correlated theories that build upon a single CSF-ROHF reference state.
One approach is to use the single CSF as a reference for multiconfigurational correlation theories.
For example, the second-order perturbative correction proposed by Song \textit{et al.}, which is based on L\"owdin downfolding\cite{Lowdin1951,Lowdin1962a} 
and exploits the stochastic split generalized active space approach,\cite{Bonfirraro2025}
can predict the correct spin-state ordering in  transition metal clusters and provide CASSCF-quality orbitals.\cite{Song2026}
These results demonstrate the importance of optimizing the CSF-ROHF orbitals in the presence of dynamic
correlation effects.

The success of a perturbative correction encourages the consideration of other
correlated CSF-based methods that may have lower computational scaling.
An obvious candidate is to combine CSF-ROHF theory for describing
spin coupling with a DFT approach for dynamic correlation, which both have mean-field scaling.
This unification has previously been achieved for open-shell singlet spin coupling  through the 
restricted open-shell Kohn--Sham (ROKS) approach.\cite{Filatov1998,Frank1998}
In this work, we aim to extend ROKS to an arbitrary genealogical spin coupling, introducing the 
CSF-ROKS approach, and assess its performance for low-spin open-shell electronic structure.

In KS-DFT, a set of orbitals is introduced to define the Slater determinant of an 
auxiliary non-interacting system, with an effective potential constructed to reproduce 
the exact one-body density.\cite{Kohn1965}
For molecules with unpaired electrons, it is standard practice to use a spin-polarized (unrestricted) 
formalism that allows the spin-up and spin-down orbitals to have different spatial 
parts.\cite{Barth1972,Rajagopal1973,Gunnarsson1976,Pople1995}
The resulting Slater determinant is generally not an eigenfunction of $\hat{S}^2$, 
but this ``spin contamination'' is formally irrelevant, since the auxiliary system 
need only reproduce the ground-state spin density.\cite{Pople1995,Perdew2009}
In practice, densities that break spin symmetry can complicate the interpretation of KS-DFT 
results as it can be difficult to determine if a given solution approximates 
the state of interest,\cite{Verma2020} particularly for open-shell systems with 
nearly degenerate spin states.

There have been several attempts at generalizing DFT to multiconfigurational states. 
A key challenge is avoiding any double counting of dynamic correlation, which is captured by both the
multiconfigurational wavefunction and the density functional approximation.
Early schemes considered additive approaches, where the energy components are partitioned into
contributions from the multiconfigurational wavefunction and the 
density functional.\cite{Lie1974a,Lie1974b,Savin1988,Grafenstein2000,Gusarov2004,Grafenstein2005,Jimenez2007}
Adaptations based on range-separated functionals aim to avoid double counting by 
splitting the electron--electron interaction such that long-range static correlation is treated by 
wavefunction theory and short-range dynamic correlation is captured by the 
functional.\cite{Savin1995,Savin1996,Leininger1997,Fromager2007,Hedegard2015,Hedegard2018,Pernal2021}
Alternatively, in multiconfigurational pair-density functional theory, only the one-electron components are 
described by wavefunction theory, while the two-electron interactions are computed from a sum of the Coulomb 
interaction and an on-top density functional.\cite{Manni2014,Gagliardi2017,Sharma2021}
However, all these methods rely on either an active space or a truncated CI wavefunction
with high computational scaling.

In contrast, the ROKS approach aims to compute energies specifically for spin multiplets
and does not rely on an active space reference.
This framework is based on the fact that the HF energy of a spin multiplet can be written 
as a weighted sum of single determinant energies.\cite{Ziegler1977}
For example, the energy of the open-shell singlet 
\begin{equation}
\ket{\Psi_\text{os}} = \frac{1}{\sqrt{2}} \qty(  |\phi_1 \bar{\phi}_2|  +  |\phi_2 \bar{\phi}_1| )
\label{eq:oss_wfn}
\end{equation}
can be written as a weighted sum of determinant energies
\begin{equation}
E[\Psi_\text{os}] = 2E\qty[ |\phi_1 \bar{\phi}_2| ] - E\qty[ |\phi_1 \phi_2| ]\\
\label{eq:oss}
\end{equation}
where the absence (presence) of an overbar denotes a high (low) spin orbital.
This approach is closely related to Yamaguchi's spin projection.\cite{Yamaguchi1988}
In 1998, both Filatov and Shaik,\cite{Filatov1998} and Frank \textit{et al.}\cite{Frank1998} 
proposed to optimize the energy for an ensemble of determinants constructed from a single set of spin-restricted orbitals, 
introducing the ROKS method.
These approaches focused on the open-shell singlet case using the ensemble defined in Eq.~\eqref{eq:oss}.
(Note that there is also a version of ROKS for single-determinant high-spin states.\cite{Murray1993,Russo1994,Okazaki1998})
Since then, ROKS has been applied almost exclusively to open-shell singlets and has proved 
particularly useful for describing excited states.\cite{Frank1998,Grimm2003,Nonnenberg2003,Kowalczyk2013,Hait2016,Shee2020,Hait2020a,Hait2021,Kunze2021,Cunha2022}
It can describe orbital relaxation in core excitations,\cite{Hait2020a,Cunha2022}
and provides accurate energies for charge transfer excitations
where the accuracy of time-dependent DFT deteriorates due to delocalization errors.\cite{Hait2016,Kunze2021} 
Limited extensions beyond the open-shell singlet have been reported,\cite{Filatov1999b,Schulte2010}
but these approaches require a separate derivation for each multiplet and have not been
 formulated for an arbitrary genealogical spin-coupling pattern.
Furthermore, the restricted ensemble KS (REKS) provides an extension  to (nearly) degenerate
configurations that are not related by spin symmetry\cite{Filatov1999,Filatov2015}
e.g., closed-shell configurations in diradicaloids, but relies on an active space 
model and has not been implemented for more than four electrons in four orbitals.\cite{Filatov2016a,Filatov2016b,Filatov2017}

These successes and limitations pose the question of whether ROKS can be extended to an arbitrary 
genealogical spin-coupled configuration.
Such an approach would enable a spin-pure prediction of low-spin open-shell states within a 
restricted KS approach, for an arbitrary number of unpaired electrons and without any active space
reference.
The central challenge is finding the form of the ensemble expansion for an arbitrary genealogical spin coupling pattern.
We desire that this method should satisfy three conditions:
(i) the ensemble should contain a polynomially-scaling number of determinants to 
avoid any large increase in computational cost;
(ii) the corresponding CSF energy should be recovered in the HF limit;
and 
(iii) the energy for spin-coupled non-interacting states should be size consistent.

In this work, we introduce the CSF-ROKS method to achieve these aims for any genealogical 
spin-coupling pattern.
The number of determinants in the ensemble scales quadratically with the 
number of open orbital shells (i.e., consecutive orbitals with the same spin coupling),
leading to a minimal increase in computational cost compared to the open-shell singlet case.
We show how the restricted orbitals can be optimized using the geometric direct minimization
algorithm for a single CSF, recently developed by one of the authors,\cite{Burton2025}
and test its performance for a range of open-shell systems.
To the best of our knowledge, these advances provide the first spin-pure extension of ROKS
to an arbitrary genealogical spin-coupling pattern, laying the foundation for spin-pure DFT 
calculations of diverse low-spin open-shell states.

\section{Background Theory}

\subsection{The ROKS formalism}
\label{sec:roks}
Following Ziegler's sum approach,\cite{Ziegler1977} the energy of a spin multiplet 
$\Psi$ can be written as the weighted sum of single determinant energies as 
\begin{equation}
E[\Psi] = \sum_{L} c_L E[\Phi_L],
\label{eq:zieglersum}
\end{equation}
where each $\Phi_L$ is an unrestricted Slater determinant with a different set of 
occupied orbitals.
The advantage of this decomposition is that the multiplet exchange-correlation energy $E_\text{xc}$ 
can be computed using a weighted sum of contributions from each Slater determinant as 
\begin{equation}
E_\text{xc}[\Psi] = \sum_{L} c_L E_\text{xc}[\Phi_L],
\label{eq:zieglersumXC}
\end{equation}
enabling the use of standard single-reference KS-DFT machinery.
The expansion coefficients are normalized to satisfy $\sum_{L} c_L = 1$ but, unlike 
 a true ensemble approach,\cite{Gould2026} the $c_L$ are not constrained to be positive weights.
Crucially, the expansion coefficients are chosen such that the ensemble energy matches 
the wavefunction energy at the HF level.
For example, the energy of the open-shell singlet defined in Eq.~\eqref{eq:oss_wfn} 
can be written exactly using the weighted sum of two single determinant energies
in Eq.~\eqref{eq:oss}.
This equivalence can be seen by comparing the energies of the spin-symmetry-broken (SB)
determinant $\ket{\Phi_\text{SB}} \equiv |\phi_1 \bar{\phi}_2| $ and the high-spin triplet (T) determinant  $\ket{\Phi_\text{T}} \equiv |\phi_1 \phi_2| $, 
given by 
\begin{align}
E\qty[\Phi_\text{SB}] &= \Vn + h_{11} + h_{22} + \expval{12|12} \\
E\qty[\Phi_\text{T} ]&= \Vn + h_{11} + h_{22} + \expval{12|12} - \expval{12|21}
\end{align}
with the open-shell singlet energy
\begin{equation}
E\qty[\Psi_\text{os}] = \Vn + h_{11} + h_{22}  + \expval{12|12} + \expval{12|21}.
\end{equation}
In this case, the  ensemble expansion coefficients  are $c_\text{SB} = 2$ and 
$c_\text{T} = -1$, recovering the formula in Eq.~\eqref{eq:oss}.

The most general form of Ziegler's sum method allows each 
determinant in the expansion to be built from a different set of molecular orbitals.
Early applications used individually optimized unrestricted KS 
solutions,\cite{Gunnarsson1976,Daul1994} but this does not guarantee that 
symmetries are correctly conserved.
Instead, the ROKS approach enforces that all the determinants in the ensemble are 
constructed from a single spin-restricted set of orbitals $\{\phi_p\}$.\cite{Filatov1998,Frank1998}
The only difference between each determinant is whether each open-shell orbital
is occupied by a spin-up or spin-down electron.
The open-shell singlet energy Eq.~\eqref{eq:oss} is then optimized with 
respect to the orbital coefficients through a modified open-shell Fock 
operator.\cite{Filatov1998}
This strategy requires only a single SCF calculation to optimize the orbitals and compute
the multiplet energy, and ensures that any desired spin symmetry properties are enforced exactly.

\subsection{Genealogical configuration state functions}
\label{sec:csf}

A single CSF is defined as an eigenfunction of the orbital occupation number operators $\hat{n}_p$
and both the projected and total spin operators.\cite{HelgakerBook}
Each CSF corresponds to a set of spatial molecular orbitals (MOs) $\{\phi_p\}$ expanded in an 
AO basis $\{\chi_\mu\}$ as
\begin{equation}
\ket{\phi_p} = \sum_\mu \ket{\chi_\mu} C_{\mu p}
\end{equation} 
and a spin-coupling pattern dictating how the unpaired electrons couple to an eigenfunction of $\hat{S}^2$.
Each MO may be occupied by zero, one, or two electrons.

In the genealogical scheme, the spin coupling is constructed by incrementally adding one 
unpaired electron at a time and coupling it to the existing electrons to maintain a spin eigenstate at 
every intermediate step.\cite{HelgakerBook}
This construction is denoted by a relative spin-coupling vector $\bm{t}$ 
that describes how the total spin changes as each electron is coupled.
For example, the first orbital corresponds to a doublet state ($\bm{t}=[{+}]$), 
and the second electron may be coupled to create either a singlet ($\bm{t}=[{+}{-}]$) or triplet state ($\bm{t}=[{+}{+}]$).
This spin-coupling pattern defines the determinantal expansion of the CSF 
\begin{equation}
\ket{\Psi(\bm{t})} = \sum_I a_I(\bm{t}) \ket{\Phi_I},
\end{equation}
where the $\ket{\Phi_I}$ are determinants with 
different assignments of up and down spins to the open-shell orbitals,
and the coefficients $a_I(\bm{t})$ are fixed by Clebsch--Gordan coupling rules.
Beyond two electrons, there may be multiple CSFs that give rise to the same total spin $S$ through
different spin-coupling patterns, e.g.,
three electrons can couple to a doublet through an intermediate triplet $[{+}{+}{-}]$ or an 
intermediate singlet $[{+}{-}{+}]$.
A CSF may also be classified according to its seniority $\Omega$, corresponding to the number
of unpaired electrons, e.g.\ $[{+}{+}{-}]$ and $[{+}{-}{+}]$ are both seniority-3 states with $\Omega=3$.

The energy of a single CSF is defined as 
\begin{equation}
E = \sum_p n_p  h_{pp} + \frac{1}{2} \sum_{pq}\qty(a_{pq} \braket{pq}{pq} + b_{pq} \braket{pq}{qp})
\end{equation}
where $h_{pq}$ denote the one-electron integrals, $\braket{pq}{rs}$ are the two-electron 
integrals, $n_p$ is the orbital occupation number, and the nuclear repulsion is omitted for brevity.
The constants $a_{pq}$ and $b_{pq}$ define the Coulomb and exchange interaction
between orbitals $\phi_p$ and $\phi_q$, respectively, and are closely related to Roothaan's 
vector coupling constants.\cite{Roothaan1960}
While $a_{pq} = n_p n_q$ is a classical Coulomb contribution, the exchange 
constants $b_{pq}$ depend on the spin-coupling pattern and are defined as\cite{Burton2025}
\begin{equation}
b_{pq} = (1-\delta_{pq})\mel{\Psi}{\hat{E}_{pq}\hat{E}_{qp}}{\Psi} - n_p,
\end{equation}
where $\hat{E}_{pq} = \sum_{\sigma \in \{\uparrow,\downarrow\}} \hat{a}^\dagger_{p\sigma} \hat{a}_{q\sigma}$ is the singlet excitation operator.\cite{Paldus1976a}
The $b_{pq}$ constants can be computed using the graphical unitary group approach (GUGA).\cite{Paldus1974,Paldus1976a,Paldus1976b,Shavitt1977}

The CSF orbitals can be subdivided into $\Ns$ open shells corresponding to orbitals that 
share the same incremental change in total spin, labeled by calligraphic indices $\cP, \cQ,$ etc.
For example, the doublet configuration $[{+}{+}{-}]$ includes the doubly occupied shell $\mathcal{C}$, and two 
sequential singly occupied shells $\cP = [{+}{+}]$ and $\cQ=[-]$.
The unoccupied orbitals can also be considered as the virtual shell $\cA$.
Crucially, the CSF energy and wavefunction are invariant to rotations between MOs within the same shell.\cite{Burton2025}
This invariance means that all open-shell orbitals $p$ in shell $\cP$ and $q$ in shell $\cQ$ share the 
same exchange constant $b_{pq}$,
which can therefore be computed once for each pair of open shells and can be
represented by the $\Ns\times \Ns$ shell exchange coupling matrix $\beta_{\cP \cQ} \equiv b_{pq}$.

\section{CSF-ROKS Theory}
\subsection{Ensemble expansion for arbitrary spin coupling}
\label{sec:csf_roks}
We now consider how to construct a suitable KS ensemble for an arbitrary genealogical CSF,
generalizing the open-shell singlet result.
Consider a general CSF with doubly-occupied orbitals $\phi_i, \phi_j, ...$, open-shell orbitals 
$\phi_p, \phi_q, ...$, unoccupied orbitals $\phi_a, \phi_b, ...$, and spin-coupling vector $\bm{t}$.
Our aim is to find an ensemble that yields the same HF energy as the CSF wavefunction.
The unrestricted determinants that comprise this ensemble all share the same total orbital occupation 
numbers, but differ in the spin occupancy of the open-shell orbitals.

Since all the determinants share the same total orbital occupation, 
they have the same one-electron energy and Coulomb repulsion and their energies only 
differ in the exchange contribution.
Therefore, to find the ensemble expansion coefficients, it is sufficient to fit the linear combination of 
individual exchange contributions for each determinant to the overall CSF exchange energy.
It is convenient to write the CSF exchange energy in the form
\begin{equation}
K[\Psi] = \frac{1}{2} \sum_{pq} b_{pq} \expval{pq|qp},
\end{equation}
and similarly for the ROKS ensemble energy
\begin{equation}
K[\Psi] =  \frac{1}{2}  \sum_{pq} \sum_{L} c_L b^{L}_{pq} \expval{pq|qp},
\end{equation}
where $b^{L}_{pq}$ are the orbital exchange coupling constants for determinant $\Phi_L$.
Mathematically, the task is now to find the set of determinants and coefficients that satisfy 
\begin{equation}
b_{pq} =  \sum_{L} c_L b^{L}_{pq} .
\label{eq:coeffSum}
\end{equation}
As discussed in Section~\ref{sec:csf}, the coefficients $b_{pq}$ are set by the genealogical spin coupling 
pattern and are the same for all $p \in \cP$ and $q \in \cQ$.
If the diagonal elements are forced to be equal ($-1$), then there are only $d = 1+ \frac{1}{2}\Ns(\Ns-1)$ independent values 
of $b_{pq}$, which correspond to the  lower triangular block of the shell exchange coupling 
matrix $\beta_{\cP \cQ}$ for the $\Ns$ singly-occupied shells.

To solve for these coefficients,
we first consider the matrix $\beta_{\cP \cQ}$ as an element of the vector space 
\begin{equation}
V = \{  \bm{M} \in \mathbb{R}^{\Ns \times \Ns} \,|\,  \bm{M} = \bm{M}^\intercal \text{ and } M_{ii} = a \},
\end{equation}
i.e., the set of symmetric matrices with a constant diagonal $a$.
This vector space has dimension $d$, accounting for the off-diagonal values in $\beta_{\cP \cQ}$ and the 
constant diagonal.
Therefore, the independent elements in $\bm{\beta}$ form a $d$-dimensional vector $\bm{v}$.
For example,  a system with four shells has the shell exchange coupling matrix $\bm{\beta} \in V$ as
\begin{equation}
\bm{\beta} = 
\begin{pmatrix}
-1 & \beta_{21} & \beta_{31} & \beta_{41}  \\
\beta_{21} & -1 & \beta_{32} & \beta_{42} \\
\beta_{31} & \beta_{32} & -1 & \beta_{43} \\
\beta_{41} & \beta_{42} & \beta_{43} & -1 
\end{pmatrix},
\end{equation}
with $a=-1$. 
The corresponding vector representation is 
\begin{equation}
\bm{v} = (-1, \beta_{21}, \beta_{31}, \beta_{41}, \beta_{32}, \beta_{42}, \beta_{43})^{\intercal}.
\end{equation}

We now need to define a basis of vectors in which to expand and represent $\bm{v}$. 
These basis vectors should translate directly to an unrestricted Slater determinant used to build the CSF ensemble.
In our construction, we only need to consider unrestricted Slater determinants where all the electrons occupying orbitals
that correspond to the same shell in the CSF share the same spin occupation.
We can then write down a shell exchange matrix $\bm{\beta}^J \in V$ for each unrestricted determinant $\Phi_J$,
 where the off-diagonal matrix elements $\beta^J_{pq}$ are $-1$ if the electrons in two shells have the same spin, and 
0 otherwise.
For our example of a 4-shell CSF, the determinant with high-spin electrons in orbitals corresponding to shells 1 and 2
and low-spin electrons in orbitals corresponding to shells 3 and 4 would have the exchange coupling matrix
\begin{equation}
\bm{\beta}^{J} = 
\begin{pmatrix}
-1 & -1& 0 & 0 \\
-1 & -1 & 0 & 0 \\
0 & 0 & -1 & -1 \\
0 & 0 & -1 & -1 
\end{pmatrix},
\end{equation}
with the corresponding vector form 
\begin{equation}
\bm{v}^{J} = (-1, -1, 0, 0, 0, 0, -1)^{\intercal}.
\end{equation}
We denote this spin arrangement as $[\alpha,\alpha,\beta,\beta]$ and note that the same matrix would be obtained 
under a complete spin reversal to the arrangement $[\beta,\beta,\alpha,\alpha]$.
Therefore, there are $2^{\Ns-1}$ different possible ways to distribute spin-up or spin-down electrons among $\Ns$ shells,
where we enforce that the first shell is occupied by spin-up electrons to avoid the spin-reversal equivalency.

Since the vector space $V$ has dimension $d = 1 + \frac{1}{2}\Ns(\Ns-1)$, we only need to find $d$ independent
basis vectors from the $2^{\Ns-1}$ possible arrangements of spins among the shell. 
Recognizing that 
\begin{equation}
1 + \frac{1}{2}\Ns(\Ns-1) = 1 + \binom{\Ns-1}{1} + \binom{\Ns-1}{2},
\end{equation}
we see that this can be achieved by including the determinant with a fully high-spin open-shell 
arrangement $[\alpha,\alpha,...]$, the $\binom{\Ns-1}{1}$ determinants obtained by every possible way 
to assign low-spin electrons to one shell (excluding the first shell), and the $\binom{\Ns-1}{2}$ determinants 
obtained by every possible way to assign low-spin electrons to two shells (excluding the first shell).
For example, with four open shells, the different spin arrangements that can be used to define the basis
for the ensemble are 
\begin{equation}
\begin{split}
&[\alpha,\alpha,\alpha,\alpha] 
\\
&[\alpha,\alpha,\alpha,\beta], \quad
[\alpha,\alpha,\beta,\alpha], \quad
[\alpha,\beta,\alpha,\alpha] 
\\
&[\alpha,\alpha,\beta,\beta], \quad 
[\alpha,\beta,\alpha,\beta], \quad 
[\alpha,\beta,\beta,\alpha] .
\end{split}
\end{equation}
Note that, for $m \ge4$, the number of possible spin arrangements is larger than the vector space, 
so there is a redundancy in how we choose the basis for the ensemble. 
The ensemble energies would be the same at the HF level for any acceptable ensemble, but 
they can change when an exchange-correlation functional is included.
In practice, we find that this choice has a small impact on numerical results.

Now that we have found which determinants $\Phi_J$ to include in the CSF-ROKS ensemble, 
we need to find their expansion coefficients $c_J$.
This can be achieved by expanding the vector form of the CSF shell exchange coupling matrix $\bm{v}$
as a linear combination of the ensemble determinant basis vectors as 
\begin{equation}
\bm{v} = \sum_{L} \bm{v}^L c_L .
\label{eq:v_exp}
\end{equation}
Note that, since $v^L_1 =  -1$ for all $L$ and $v_1 =-1$, the constraint $1 = \sum_L c_L$ is encoded within this expansion.
Taking the dot product of Eq.~\eqref{eq:v_exp} with the basis vector $\bm{v}^J$ gives the linear equation
\begin{equation}
\expval*{\bm{v}^J \cdot \bm{v} } = \sum_{L} g^{JL} \, c_L,
\label{eq:v_linprob} 
\end{equation}
where $g^{JL} = \expval*{\bm{v}^{J} \cdot \bm{v}^L}$ is the metric tensor of the basis vectors,
which are linearly independent but not necessarily orthogonal.
It is then straightforward to obtain the expansion coefficients $c_L$ by solving the linear problem Eq.~\eqref{eq:v_linprob}.

To illustrate this approach, we construct the ensemble expansion for the open-shell triplet CSF with the spin coupling pattern
$\bm{t} = [{+}{+}{-}{-}{+}{+}]$, which contains three open shells.
The shell exchange coupling matrix can be computed using the GUGA framework to give
\begin{equation}
\bm{\beta} = 
\begin{pmatrix}
-1 & \frac{1}{2} & - \frac{1}{2} \\
\frac{1}{2} &  -1  & -\frac{1}{2} \\
-\frac{1}{2} & -\frac{1}{2} &  -1 
\end{pmatrix}
\end{equation}
with the vectorized form
\begin{equation}
\bm{v} = (-1, \frac{1}{2},-\frac{1}{2},-\frac{1}{2})^\intercal.
\end{equation}
We introduce four basis determinants with the following arrangements of high- and low-spin electrons in the three open shells 
as $[\alpha,\alpha,\alpha]$, $[\alpha,\alpha,\beta]$, $[\alpha,\beta,\alpha]$, and $[\alpha,\beta,\beta]$,
giving rise to determinants with the spin occupancies
\begin{equation}
\begin{split}
[\alpha,\alpha,\alpha]&: \qquad \Phi_1 = |\phi_1 \phi_2 \phi_3 \phi_4 \phi_5 \phi_6|,\\
[\alpha,\alpha,\beta]&: \qquad \Phi_2 = |\phi_1 \phi_2 \phi_3 \phi_4 \bar{\phi}_5 \bar{\phi}_6|,\\
[\alpha,\beta,\alpha]&: \qquad \Phi_3 = |\phi_1 \phi_2 \bar{\phi}_3 \bar{\phi}_4 \phi_5 \phi_6|,\\
[\alpha,\beta,\beta]&: \qquad \Phi_4 = |\phi_1 \phi_2 \bar{\phi}_3 \bar{\phi}_4 \bar{\phi}_5 \bar{\phi}_6|.\\
\end{split}
\end{equation}
The corresponding basis vectors are
\begin{equation}
\begin{split}
[\alpha,\alpha,\alpha]&: \qquad \bm{v}^{1} = (-1,-1,-1,-1)^\intercal, \\
[\alpha,\alpha,\beta]&: \qquad \bm{v}^{2} = (-1,-1,0,0)^\intercal, \\
[\alpha,\beta,\alpha]&: \qquad \bm{v}^{3} = (-1,0,-1,0)^\intercal, \\
[\alpha,\beta,\beta]&: \qquad \bm{v}^{4} = (-1,0,0,-1)^\intercal .
\end{split}
\end{equation}
Solving Eq.~\eqref{eq:v_linprob} then leads to the ensemble expansion
\begin{equation}
E[\Psi] = -\frac{1}{4} \qty(E[\Phi_1] + E[\Phi_2] - 3 E[\Phi_3] - 3 E[\Phi_4]).
\end{equation}

Here, we have shown how an ensemble expansion of unrestricted Slater determinants 
can be constructed to exactly recover the HF energy of an arbitrary CSF constructed through genealogical 
spin coupling.
This construction takes advantage of recently derived expressions for the CSF shell exchange matrix elements 
$\beta_{\cP \cQ}$ using the GUGA framework.\cite{Gouveia2024,Burton2025}
Crucially, the number of determinants in this expansion scales with the number of open-shells $\Ns$ as 
$\mathcal{O}(\Ns^2)$, even though the number of determinants 
in a CSF expansion grows at worst combinatorially with the number of singly-occupied orbitals.
This scaling ensures that the computational cost of CSF-ROKS also scales quadratically with the number of open shells.

\subsection{Optimizing the CSF-ROKS orbitals}
\label{sec:csf_roks_opt}


We now consider how to optimize the CSF-ROKS orbitals using a self-consistent field (SCF) theory.
Our approach follows the CSF-based geometric direct minimization (CSF-GDM) procedure introduced in Ref.~\citenum{Burton2025},
which provides robust quasi-Newton energy minimization and avoids issues related to defining a universal 
Fock operator in ROHF theory.\cite{Roothaan1960,Segal1970,Peters1972,Binkley1974,Guest1974,Edwards1987,Tsuchimochi2010,Gouveia2024}
Since our optimization algorithm closely follows Ref.~\citenum{Burton2025}, here we will only highlight the 
most important details and any modifications required for CSF-ROKS theory.

In the CSF-GDM approach, we require a local parametrization for steps on the orbital
constraint surface, corresponding gradient elements, and an approximation to the 
diagonal of the orbital rotation Hessian to accelerate convergence using preconditioning.
The CSF orbital coefficients on the $k$th step are updated according to an 
exponential transformation with the form 
\begin{equation}
\bm{C}^{(k+1)} = \bm{C}^{(k)} \exp(\bm{\kappa})
\label{eq:expK}
\end{equation}
where $\bm{\kappa}$ is an anti-Hermitian matrix of orbital rotation parameters.
The $\bm{\kappa}$ step matrix only includes non-zero entries for rotations between orbitals in different shells, 
e.g., a CSF with four shells (including core, two open shells, and virtual) has steps with the form
\begin{equation}
\bm{\kappa} = 
\begin{pmatrix}
\bm{0} & -\bm{\kappa}_{21}^\intercal & -\bm{\kappa}_{31}^\intercal & -\bm{\kappa}_{41}^\intercal
\\
\bm{\kappa}_{21} & \bm{0} & -\bm{\kappa}_{32} ^\intercal & -\bm{\kappa}_{42}^\intercal
\\
\bm{\kappa}_{31} & \bm{\kappa}_{32} & \bm{0} & - \bm{\kappa}_{43}^\intercal
\\
\bm{\kappa}_{41} & \bm{\kappa}_{42} & \bm{\kappa}_{43} & \bm{0}
\end{pmatrix}.
\end{equation}
The mathematical structure of this invariance to intra-shell rotations and the resulting 
parametrization of the CSF orbital coefficients corresponds to a flag manifold.\cite{Vidal2024,Burton2025}

To derive the gradient of the CSF-ROKS energy with respect to these orbital rotation parameters,
we follow a generalized Kohn--Sham (GKS) procedure\cite{Seidl1996,Gould2021} to obtain
\begin{equation}
g_{pq} \equiv \frac{\partial E}{\partial \kappa_{pq}} = 2(F_{qp}^\cQ - F_{pq}^\cP),
\label{eq:CSFROKSgrad} 
\end{equation}
where $F_{pq}^\cP$ is an element of a generalized Fock operator
\begin{equation}
F_{pq}^{\cP} = n_\cP(h_{pq} + J_{pq}) + \xi K^{\cP}_{pq} +  \mel{\phi_p}{\hat{v}_\text{xc}^{\cP}}{\phi_q}.
\label{eq:gFock}
\end{equation}
Here, $n_\cP$ is the orbital occupation for shell $\cP$, the Coulomb and exchange matrix elements are
\begin{subequations}
\begin{align}
J_{pq}& =  \sum_{r} n_r \braket{pr}{qr},
\\
K_{pq}^\cP& =  \sum_{\cR} \beta_{\cP \cR} \sum_{r\in \cR} \braket{pr}{rq},
\end{align}
\end{subequations}
and the exact HF exchange term in Eq.~\eqref{eq:gFock} is scaled by its contribution $\xi$ in a hybrid exchange-correlation functional.
The operator $\hat{v}_\text{xc}^{\cP}$ is a shell-dependent effective exchange-correlation potential 
(excluding any HF exchange) defined using the ensemble of unrestricted determinants as
\begin{equation}
\hat{v}_\text{xc}^{\cP} = \sum_{L}\sum_{\sigma \in \{\uparrow \downarrow\}} c_L n_{\cP\sigma}^L \hat{v}_{\text{xc},\sigma}^{L},
\end{equation}
where $c_L$ are the ensemble expansion coefficients, $n_{\cP\sigma}^L$ is the 
occupation of spin-$\sigma$ particles in shell $\cP$ for the ensemble determinant $\Phi_L$,
and $\hat{v}_{\text{xc},\sigma}^{L}$ is the standard GKS spin-$\sigma$ exchange-correlation potential operator for determinant $\Phi_L$.
No assumption is made about the form of exchange-correlation functional, which may 
include non-local contributions.
A full derivation of this gradient expression is provided in Appendix~\ref{sec:gradient}.
Crucially, like the original ROKS formalism,\cite{Filatov1998,Frank1998} both the energy
and gradient can be readily computed using established routines for unrestricted KS theory. 

Finally, the convergence of quasi-Newton optimization schemes can be significantly improved by using a 
preconditioner based on approximations to the diagonal elements of the Hessian
\begin{equation}
Q_{pq,pq} \equiv \frac{\partial^2 E}{\partial \kappa_{pq} \partial \kappa_{pq}}.
\label{eq:hess}
\end{equation}
We follow the preconditioning approach described for CSF-GDM in Ref.~\onlinecite{Burton2025}.
The diagonal Hessian elements for the CSF-ROKS energy are approximated as
\begin{equation}
\begin{split}
Q_{pq,pq} \approx\ &2(F^\cP_{qq} - F^{\cQ}_{qq}) + 2(F^\cQ_{pp} - F^{\cP}_{pp})
\\
&+4(a_{pp} + a_{qq} - 2 a_{pq})\braket{pq}{qp}
\\
&+2\xi(b_{pp} + b_{qq} - 2b_{pq})\qty(\braket{pq}{qp} + \braket{pq}{pq})
\end{split}
\label{eq:prec}
\end{equation}
where we have ignored contributions from the second functional derivative of the exchange-correlation
energy (see Appendix~\ref{sec:hessian}).
The last line in Eq.~\eqref{eq:prec} arises from any exact HF exchange contributions, scaled by their contribution
$\xi$ to the exchange-correlation functional.
Following Ref.~\onlinecite{Burton2025}, we further approximate the diagonal Hessian terms for core--virtual 
rotations as 
\begin{equation}
Q_{ia,ia} \approx\ 2 (F^\cC_{aa} - F^\cA_{aa}) + 2(F_{ii}^\cA - F_{ii}^\cC).
\label{eq:prec2}
\end{equation}

\section{Computational Details}

We have implemented a pilot version of CSF-ROKS in the \textsc{Quantel} software package
developed by our group.\cite{Quantel}
This implementation uses \textsc{PySCF}\cite{pyscf} to evaluate one- and two-electron integrals, 
and the exchange--correlation energies and potentials for each ensemble determinant.
The default numerical integration grids in \textsc{PySCF} were used in all cases.

Orbital optimization primarily used the CSF-GDM algorithm from Ref.~\onlinecite{Burton2025} with the gradient
in Eq.~\eqref{eq:CSFROKSgrad} and the diagonal preconditioner in Eqs.~\eqref{eq:prec} and \eqref{eq:prec2}.
Unless otherwise stated, CSF-ROKS calculations were initialized from the converged CSF-ROHF orbitals
for the same spin-coupling pattern, which were in turn obtained using the localization initial guess procedure
described in Ref.~\onlinecite{Burton2025}.
We have also implemented a CSF-ROKS version of the limited-memory symmetric rank-1 (L-SR1) algorithm 
described in Ref.~\onlinecite{Levi2020}, which was used to target saddle points whenever CSF-GDM converged
to an undesired symmetry-broken minimum or if we want to target an excited state-specific solution.
Calculations were considered converged when the root-mean-squared value of the gradient fell below $10^{-6}$.
CASSCF and NEVPT2 calculations for the molecular ruby were performed using \textsc{PySCF}.

\begin{figure*}[htb]
	\includegraphics[width=\linewidth]{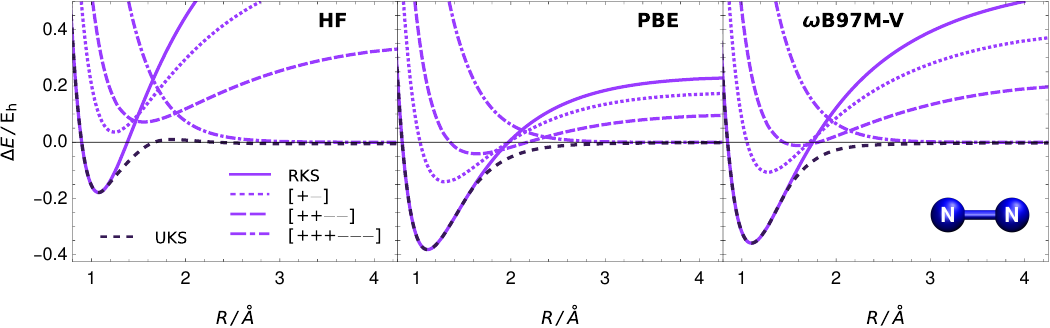}
	\caption{CSF-ROKS energies for \ce{N2} using the HF (left), PBE (middle), or $\omega$B97M-V (right)
		functionals for different spin-coupling patterns with the cc-pVDZ basis. 
		Each spin-coupling pattern is initialized using localized orbitals that reflect the partially-bond-broken configurations described previously.
		Energies are plotted relative to twice the high-spin quartet ROKS energy of a \ce{N} atom, showing that the $[{+}{+}{+}{-}{-}{-}]$ configuration dissociates to the correct limit.}
	\label{fig:n2_curve}
\end{figure*}

The cc-pVDZ basis set\cite{Dunning1989,Woon1993,Balabanov2005,Balabanov2006} was used for \ce{N2}, 
the polyacenes, and the transition-metal dimers.
Geometries for the polyacenes and bridged transition-metal dimers were taken from 
Ref.~\onlinecite{Burton2025} and Ref.~\onlinecite{Schurkus2020}, respectively.
For the vinyl and allyl radicals, we used the aug-cc-pVDZ basis set that includes 
the diffuse functions required to describe the 3s and 3p Rydberg states, using 
molecular geometries taken from the QUEST database.\cite{Loos2020}
All-electron calculations on the \ce{[Cr(ddpd)_2]^{3+}} complex were performed using the 
def2-TZVP basis set\cite{Weigend2005} with the ground state structure taken from Ref.~\onlinecite{Otto2015}.
To accelerate calculations on this chromium complex, we employed density fitting 
using the def2-TZVP/JK auxiliary basis set,\cite{Weigend2008} as implemented in \textsc{PySCF}.
All molecular geometries are available in a public repository at \href{https://doi.org/10.5281/zenodo.24438513}{10.5281/zenodo.24438513}.

\section{Results}

\subsection{Molecular \ce{N2} dissociation}
\label{sec:n2_dissoc}

We first consider the dissociation of \ce{N2}, where breaking the 
triple bond creates six strongly correlated electrons.
This example has previously been used
to demonstrate how states with many unpaired electrons can be represented using a single CSF.\cite{Manni2020,MartiDafcik2024} 
At stretched bond lengths, the ground state can 
be represented by a single CSF corresponding to two high-spin quartets (each localized on 
one \ce{N} atom) coupled antiferromagnetically to an overall singlet with the 
spin-coupling pattern $[{+}{+}{+}{-}{-}{-}]$,
whereas the equivalent (6e, 6o) active space description would require up to 400 determinants.
Ref.~\onlinecite{MartiDafcik2024} also investigated the relative energy surfaces of the closed-shell, 
$[{+}{-}]$, $[{+}{+}{-}{-}]$, and $[{+}{+}{+}{-}{-}{-}]$ spin-coupling patterns, showing that CSFs with zero, two, or four unpaired
electrons can represent fully-bonded or partially-broken bonds.
For example, a CSF with spin coupling $[{+}{+}{-}{-}]$ and open-shell orbitals corresponding to localized
$2\mathrm{p_x}$ and $2\mathrm{p_y}$ AOs represents a partial bonding configuration 
with two broken $\uppi$ bonds, where local high-spin triplets on each \ce{N} couple antiferromagnetically to 
an overall singlet state.

This physically intuitive picture of chemical bonding presents an opportunity 
to test whether CSF-ROKS provides a meaningful extension of CSF-ROHF theory.
Here, we extend Ref.~\onlinecite{MartiDafcik2024} by using the larger cc-pVDZ basis set, 
including orbital optimization, and comparing the CSF-ROHF and CSF-ROKS descriptions.
We have identified all the partially-bonded configurations using CSF-ROHF theory 
(left panel in Fig.~\ref{fig:n2_curve}) that were described previously.
Here the open-shell orbitals in the $[{+}{-}]$ CSF correspond to $2\mathrm{p_x}$ (or  $2\mathrm{p_y}$) 
orbitals on both \ce{N}, representing a single broken $\uppi$-bond. 
While Ref.~\citenum{MartiDafcik2024} considered a symmetric linear combination
of $\uppi_\text{x}$- and $\uppi_\text{y}$-broken states, here we report only the single $[{+}{-}]$ CSF energies.
As reported before, we see that increasing the number of unpaired electrons progressively breaks 
the triple bond, with the corresponding binding curves reflecting the reduction in bonding character.

Using the optimal CSF-ROHF orbitals as initial guesses, we have located the corresponding 
CSF-ROKS solution for each spin-coupling pattern.
We considered the PBE\cite{Perdew1996} and $\omega$B97M-V\cite{Mardirossian2016a} exchange-correlation functionals, 
which provide representative examples of a modern non-empirical generalized gradient approximation (GGA) functional
and a range-separated hybrid meta-GGA functional, respectively.
The resulting CSF-ROKS energy surfaces share all the qualitative partial-bonding features of the CSF-ROHF solutions 
(middle and right panels in Fig.~\ref{fig:n2_curve}), including energy minima at progressively higher energies and longer 
bond lengths for more unpaired electrons.
In all cases, the $[{+}{+}{+}{-}{-}{-}]$ spin-coupled solution dissociates to the correct limit, 
represented by twice the high-spin ROKS energy of a single \ce{N} atom with $m_s = \tfrac{3}{2}$.
Therefore, CSF-ROKS provides a chemically meaningful density functional extension of CSF-ROHF theory.

\begin{figure*}[htb!]
	\includegraphics[width=\linewidth]{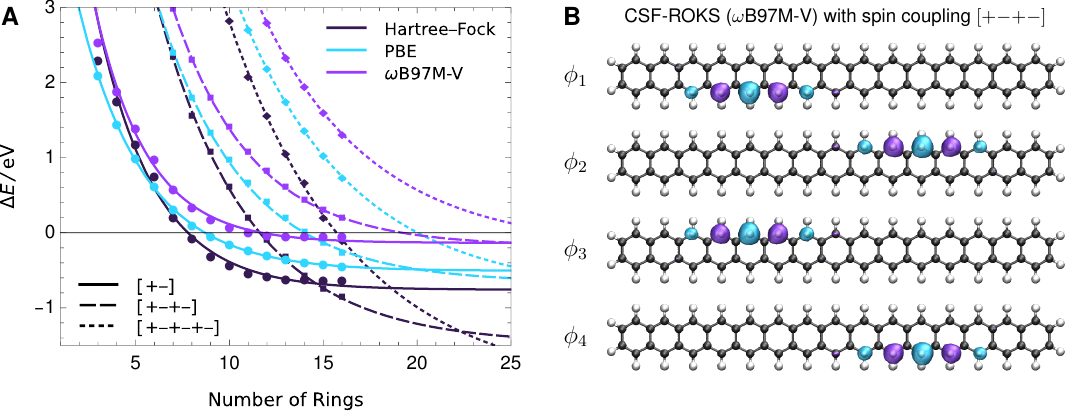}
	\caption{(\textsf{\textbf{A}}) Energies of CSF-ROKS solutions for different spin-coupling patterns shown
		relative to the corresponding closed-shell RHF/RKS solution for different functionals (cc-pVDZ).
		Including correlation through an exchange-correlation functional delays the onset of polyradical character
		to larger polyacene chains.
		(\textsf{\textbf{B}}) Optimized open-shell CSF-ROKS orbitals for the $[{+}{-}{+}{-}]$ spin-coupling pattern 
		in the orbital ordering $\phi_1$ -- $\phi_2$ -- $\phi_3$ -- $\phi_4$, showing the intrinsic localization of 
		unpaired electrons onto distinct quadrants of the 15-acene chain.}
	\label{fig:polyacene}
\end{figure*}

These \ce{N2} binding curves also allow us to understand how including an exchange-correlation 
functional differentially affects different spin-coupling patterns.
For both the PBE and $\omega$B97M-V functionals, there is relatively more energetic stabilization 
for spin-coupling patterns with fewer unpaired electrons, meaning that the closed-shell (RHF/RKS) 
configuration is lowered in energy the most relative to the dissociation limit (Fig.~\ref{fig:n2_curve}). 
This result reflects that dynamic electron correlation effects are generally larger for closed-shell states.
Furthermore, it is interesting to compare the CSF-ROKS energies with the lowest-energy 
unrestricted Kohn--Sham (UKS) approximation, which undergoes spin-symmetry breaking as the bond stretches.
In Fig.~\ref{fig:n2_curve}, we see that the UKS ground state always lies equal to or lower in energy than the  
lowest CSF-ROKS solution. 
Moving from CSF-ROHF to CSF-ROKS, the lower relative energies for low-seniority spin coupling 
shift the crossing point with the $[{+}{+}{+}{-}{-}{-}]$ solution to larger bond lengths, which appears to correlate with the 
delayed onset of spin-symmetry breaking in UKS and the disappearance of the unphysical UHF maximum 
at $R\approx\SI{1.85}{\angstrom}$.

Although no single CSF-ROKS solution can provide a complete binding curve for the dissociation of \ce{N2},
these results show that the intermediate and stretched bond lengths can be represented using a spin-pure
formalism of DFT.
Furthermore, our results indicate that CSF-ROKS with approximate functionals retains the physical 
interpretability and diabatic character observed in CSF-ROHF.\cite{MartiDafcik2024,Burton2025}
Finally, although Ref.~\citenum{MartiDafcik2024} constructed a smooth ground-state binding curve using
a nonorthogonal configuration interaction of individual CSF states, this approach is not currently 
possible for CSF-ROKS solutions as the off-diagonal Hamiltonian coupling between multiple KS-DFT solutions 
is not easily accessible.

\subsection{Polyradical character in the polyacene series}
\label{sec:polyacenes}

Previously, we exploited the diabatic nature of CSF-ROHF solutions to probe the
ground-state polyradical character in polyacene chains by comparing the relative
energy of CSF-ROHF solutions with an increasing number of unpaired electrons.\cite{Burton2025}
This analysis  indicated that the ground-state transitions from closed-shell to 
diradical $[{+}{-}]$ and then tetraradical $[{+}{-}{+}{-}]$ character as the number of
fused benzene rings increases.
However, the  polyradical character of long polyacenes remains controversial.
DMRG calculations using the $\uppi$-orbital active space indicate that the number of unpaired
electrons increases progressively for larger chains,\cite{Hachmann2007} 
which is supported by our previous CSF-ROHF analysis\cite{Burton2025} as well as the results 
of symmetry-broken DFT.\cite{Trinquier2018}
On the other hand, coupled cluster valence bond (CCVB) calculations that correlate all valence orbitals
suggest that the number of unpaired electrons does not increase as rapidly as predicted by DMRG, 
indicating that dynamic correlation may suppress polyradical character.\cite{Lee2017}
Resolving this controversy remains difficult due to the size of longer polyacenes.

Here, we revisit this question by using CSF-ROKS to introduce dynamic correlation
to the CSF-ROHF formalism.
We use the orbital optimized CSF-ROHF solutions for the closed-shell, diradical $[{+}{-}]$, 
tetraradical $[{+}{-}{+}{-}]$, and hexaradical $[{+}{-}{+}{-}{+}{-}]$ spin-coupled CSFs 
as the initial guess for CSF-ROKS calculations using the PBE and $\omega$B97M-V functionals 
in the cc-pVDZ basis set.
The largest of these calculations (hexaradical in 16-acene) included 6 open shells, 
1104 molecular orbitals, and nearly $2\times10^5$ optimization parameters, demonstrating the scalability of our approach.

The energy of each spin-coupled CSF-ROKS solution
is shown relative to the corresponding RKS energy for each polyacene in Fig.~\ref{fig:polyacene}.
The optimal open-shell $\omega$B97M-V orbitals for 15-acene with the spin-coupling pattern
$[{+}{-}{+}{-}]$ show the same localization onto subunits seen in the CSF-ROHF description (Fig.~\ref{fig:polyacene}\textcolor{blue}{B}),
and thus these CSF-ROKS solutions describe the same diabatic character as before.
Using the HF approximation, the lowest energy spin-coupling pattern transitions from 
closed-shell to diradical $[{+}{-}]$ at around 7-acene, and then to tetraradical $[{+}{-}{+}{-}]$
at around 14-acene, as reported previously.\cite{Burton2025}
In contrast, after introducing an exchange-correlation functional, the energy of 
configurations with more unpaired electrons is raised relative to the closed-shell solution 
(Fig.~\ref{fig:polyacene}\textcolor{blue}{A}).
The result is that the transition from closed-shell to diradical and then tetraradical character
shifts to longer acene chains in CSF-ROKS compared to CSF-ROHF theory, suggesting that
the inclusion of dynamic correlation delays the onset of polyradical character.

The relative stabilization of closed-shell configurations is stronger using the $\omega$B97M-V
functional compared to PBE.
Since $\omega$B97M-V is a highly-optimized rung 4  functional compared to the 
non-empirical GGA rung~2 PBE functional, we consider $\omega$B97M-V to 
provide a more accurate prediction of the dynamic electron  correlation.
Remarkably, the relative stabilization of low-seniority configurations in $\omega$B97M-V is 
so strong that diradical character appears to be only slightly preferred beyond 12-acene, 
but the $[{+}{-}]$ configuration remains less than $\SI{0.05}{\eV}$ below the RKS results for 16-acene. 
Extrapolating our data to increasingly large acene chains suggests that the closed-shell, diradical, 
tetraradical, and hexaradical configurations all become nearly degenerate in the large chain limit.
This result suggests that in the presence of dynamic correlation, the large polyacenes develop 
very limited polyradical character, supporting the CCVB analysis in Ref.~\onlinecite{Lee2017}.

\begin{table*}[htb!]
\caption{Vertical CSF-ROKS excitation energies (eV) for the low-lying excited states of the vinyl and allyl monoradicals across various functionals. 
Seniority-3 ($\Omega=3$) configurations for the vinyl and allyl radical valence excitations use the HOMO--LUMO--SOMO orbital ordering unless otherwise stated in the text. The mean-unsigned error (MUE) is computed relative to CIPSI benchmark results for the vinyl radical, and relative to theoretical best estimates (TBE) from the QUEST database for the allyl radical.\cite{Veril2021}
\textcolor{blue}{\textdagger} TBE values for the  $\mathrm{2\,^2A_2}$ and $\mathrm{3\,^2A_2}$ states are UHF-EOM-CCSD results using the aug-cc-pVDZ basis set taken from Ref.~\onlinecite{Mach2010}.}
\label{tab:organic_radical_energies}
\begin{ruledtabular}
	\begin{tabular}{llcrrrrrr}
		\textbf{Vinyl}     &  State                        &Spin Coupling &HF  &LDA &PBE &PBE0&$\omega$B97M-V&CIPSI\\ \hline
		$\mathrm{1\,^2A''}$& $\mathrm{1a''\rightarrow\ 1a'}$ &$+$         &2.86&3.52&3.37&3.26&3.12          &3.33\\
		$\mathrm{2\,^2A''}$& $\mathrm{1a' \rightarrow\ 2a''}$ &$+$        &4.67&4.15&4.28&4.40&4.29          &4.79\\
		$\mathrm{2\,^2A'}$ & $\mathrm{1a'' \rightarrow\ 2a''}$&${+}{+}{-}$&4.80&5.58&5.37&5.25&5.61          &5.65\\
		$\mathrm{3\,^2A'}$ & $\mathrm{1a'  \rightarrow\ 3s}$  &$+$        &6.12&5.26&5.98&6.22&6.19          &6.14\\ 
		$\mathrm{4\,^2A'}$ & $\mathrm{1a'  \rightarrow\ 3p_x}$&$+$        &6.73&5.69&---   &6.63&6.72        &6.73\\
        $\mathrm{5\,^2A'}$ & $\mathrm{1a' \rightarrow\ 3p_y}$&$+$         &6.93&5.97&6.66&6.93&7.07          &7.02\\
        $\mathrm{3\,^2A''}$ & $\mathrm{1a'' \rightarrow\ 3s}$&${+}{+}{-}$ &6.11&6.69&7.37&7.28&7.40          &7.46\\ 
        $\mathrm{4\,^2A''}$ & $\mathrm{1a'' \rightarrow\ 3s}$&${+}{-}{+}$ &6.86&7.43&7.82&7.78&8.08          &7.97\\
        $\mathrm{6\,^2A'}$ & $\mathrm{1a'' \rightarrow\ 2a''}$&${+}{-}{+}$&7.91&6.77&6.85&7.39&7.54          &7.47\\ 
        $\mathrm{1\,^4A'}$ & $\mathrm{1a'' \rightarrow\ 2a''}$&${+}{+}{+}$&3.37&4.27&4.54&4.33&4.52          &4.38 \\
		 \hline     
		MUE & all states                       &                                          &0.55&0.59&0.26&0.16&0.13          & --- \\ 
		MUE &$S=\frac{1}{2}, \Omega =3 $         &                        &0.94&0.52&0.29&0.21&0.07          & --- \\  
            \hline 
		\textbf{Allyl}         &  State                         &Spin Coupling  &HF  &LDA &PBE&PBE0&$\omega$B97M-V&TBE \\ \hline 
		$\mathrm{1\,^{2}B_{1}}$&$\mathrm{1b_1\rightarrow\ 1a_2}$&$+$            &4.79&3.94&3.94&4.18&4.37         &3.42\\
		$\mathrm{2\,^{2}B_{1}}$&$\mathrm{1a_2\rightarrow\ 2b_1}$&$+$            &4.94&4.11&4.27&4.56&4.74         &5.73\\
		$\mathrm{1\,^{2}A_{1}}$&$\mathrm{1a_2\rightarrow\ 3s}$  &$+$              &4.28&4.15&4.75&4.87&4.94         &4.95\\
		$\mathrm{2\,^{2}A_{1}}$&$\mathrm{1a_2\rightarrow\ 3p_z}$&$+$            &4.82&4.57&5.12&5.28&5.48         &5.50\\
		$\mathrm{1\,^{2}B_{2}}$&$\mathrm{1a_2\rightarrow\ 3p_y}$&$+$            &4.88&4.69&5.27&5.42&5.59         &5.60\\
		$\mathrm{2\,^{2}A_{2}}$&$\mathrm{1a_2\rightarrow\ 3d_{xy}}$&$+$         &6.96&6.42&6.91&7.12&7.49         &7.58\textcolor{blue}{\textsuperscript{\textdagger}} \\
		$\mathrm{3\,^{2}A_{2}}$&$\mathrm{1b_1\rightarrow\ 3p_x}$&${+}{+}{-}$    &7.61&---&8.82&8.73&9.31         &9.02\textcolor{blue}{\textsuperscript{\textdagger}} \\
		$\mathrm{4\,^{2}A_{2}}$&$\mathrm{1b_1\rightarrow\ 2b_1}$&${+}{+}{-}$    &9.68&7.71&7.83&8.57&9.12         & ---  \\
		$\mathrm{5\,^{2}A_{2}}$&$\mathrm{1b_1\rightarrow\ 2b_1}$&${+}{-}{+}$    &9.40&7.93&8.12&8.66&8.80         & ---  \\
		$\mathrm{1\,^{4}A_{2}}$&$\mathrm{1b_1\rightarrow\ 2b_1}$&${+}{+}{+}$    &4.53&5.92&6.06&5.75&5.95       &6.01\\ \hline 
		MUE& all states             &               &0.97&0.86&0.48&0.43&0.30         & ---\\
		MUE&$\Omega =1$, Rydberg &                  &0.67&0.95&0.39&0.24&0.03         & --- \\  
		MUE&$S=\frac{1}{2}, \Omega =3, (\mathrm{3\,^2A_2})$ &&1.40&---&0.20&0.29&0.29         & --- \\  
	\end{tabular}
\end{ruledtabular}
\end{table*}

While CSF-ROKS provides an intuitive picture of the open-shell singlet character,
this analysis is limited by the restriction of each calculation to a single spin-coupling pattern.
A more accurate description might be achievable by using a linear combination of different
spin-coupling patterns, either through a generalized ensemble approach or a form of DFT-CI,
and this must be a priority for future research.
However, a key advantage of the existing approach over methods such as symmetry-broken 
UKS theory is that the spin-purity of CSF-ROKS removes any ambiguity about 
which spin state is being described.

\subsection{Excited states of organic radicals}

\begin{figure}[htb!]
\includegraphics[width=\linewidth]{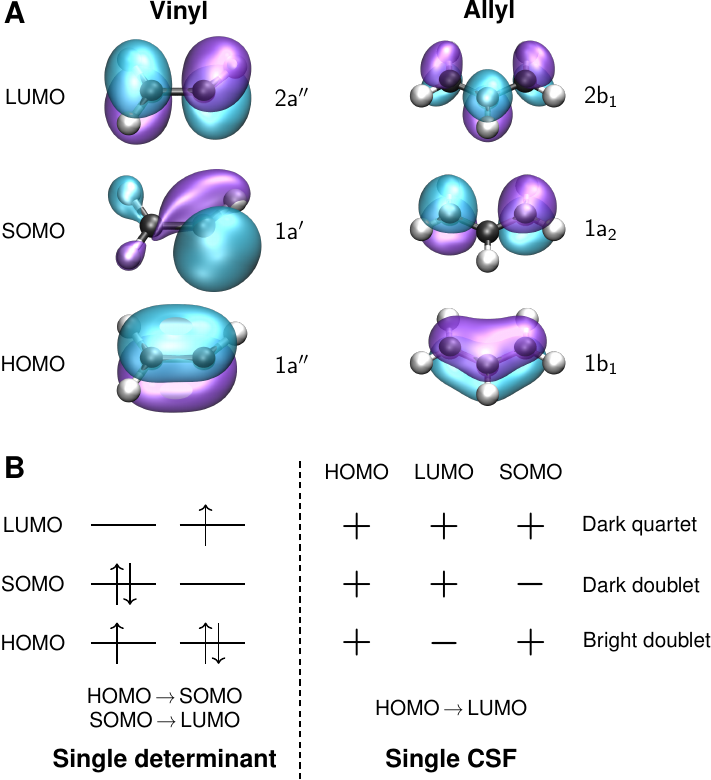}
\caption{(\textbf{\textsf{A}}) Frontier orbitals involved in the low-lying excitations for the vinyl and allyl radicals, obtained from a ground-state high-spin ROHF calculation.
(\textbf{\textsf{B}}) Classification of excitations involving the HOMO, SOMO, and LUMO in terms of single determinant or single CSF character.}
\label{fig:radicals}
\end{figure}

Simulating excited states of radicals with low-scaling methods remains a challenge due 
to spin contamination in truncated CI.\cite{Green2024} 
CSF-ROKS thus offers a potential route to spin-pure correlated excited-state approximations via a state-specific $\Delta$SCF approach.\cite{Gilbert2008,Hait2016,Burton2022}
Here, we investigate its performance on the low-lying excited states of the vinyl and allyl monoradicals. We consider the
low-lying excitations governed by transitions between the valence frontier orbitals, including the highest occupied molecular orbital (HOMO), the singly occupied molecular orbital (SOMO), the lowest unoccupied molecular orbital (LUMO) and Rydberg $n=3$ orbitals.\cite{Koziol2006,Loos2020} 
Single SOMO \textrightarrow\ LUMO or HOMO \textrightarrow\ SOMO excitations generate spin-pure seniority-1 doublets corresponding to a single Slater determinant (Fig.~\ref{fig:radicals}). 
In contrast, excitations from HOMO to virtual orbitals (including the LUMO and Rydberg orbitals) 
lead to three singly occupied orbitals, requiring the inclusion of both single and double spin-orbital excitations to maintain spin purity.\cite{Green2024} 
As shown in Fig.~\ref{fig:radicals}, this excitation space is spanned by one quartet and two doublet CSFs, 
where the first two singly occupied orbitals (e.g., the HOMO and LUMO) are either singlet-coupled in the bright $[{+}{-}{+}]$ 
state, or triplet-coupled in the dark $[{+}{+}{-}]$ state.\cite{Zhang2003,Green2024} 
Crucially, while the seniority-1 and quartet seniority-3 states can be described by standard high-spin ROKS, 
orbital optimization of the doublet seniority-3 states requires the more general CSF-ROKS approach.

For the vinyl radical, CSF-ROKS generally yields more accurate vertical excitation energies than CSF-ROHF across the range
of states and functionals studied (Table~\ref{tab:organic_radical_energies}). 
The mean unsigned error (MUE) with respect to spin-pure benchmark values computed using CIPSI\cite{QuantumPackage2.0} 
decreases as the functionals ascend the DFT functional ladder, with the most accurate excitation energies achieved 
using the $\omega$B97M-V functional (MUE of \SI{0.13}{\eV}). 

The same trend is observed for the seniority-3 singlet- and triplet-coupled doublet excited states, with $\omega$B97M-V energies 
in good agreement to the CIPSI reference (MUE of \SI{0.07}{\eV}).
In general, the singlet-coupled doublet $6\,\mathrm{^2A'}$ state  is higher in energy than the triplet-coupled counterpart $2\,\mathrm{^2A'}$, 
reflecting the larger exchange interaction between the HOMO and LUMO. 
We additionally report excitations of HOMO \textrightarrow\ $3s$ character. In this case, the same energetic ordering of the triplet and singlet coupled states is observed, but the diffuse nature of the Rydberg orbital creates a small exchange integral with the valence orbitals, 
meaning that the $[{+}{+}{-}]$ spin coupling with the HOMO -- SOMO -- Rydberg orbital ordering is preferred 
to maximize the HOMO--SOMO exchange integral.
We note that the singlet-coupled $6\,\mathrm{^2A'}$  CIPSI reference state contains significant contributions from 
high-energy excited configurations, which may represent orbital relaxation or valence-Rydberg mixing. 

Both CSF-ROHF and CSF-ROKS predict the triplet-coupled $2\,\mathrm{^2A'}$ state to be dark, with a transition dipole moment of zero, and the singlet-coupled $6\,\mathrm{^2A'}$  to be bright, in qualitative agreement with previous results
and benchmark CIPSI results (Table~\ref{tab:organic_radical_osc_strength}). 
We note that the CSF-ROKS oscillator strengths do not show high quantitative accuracy compared to CIPSI, suggesting that the
corresponding transition dipole moments are not well described by the auxiliary CSF wavefunction.

\begin{table}[hb!]
\caption{Oscillator strengths for vinyl $\mathrm{1a'' \rightarrow\ 2a''}$ excitations. CIPSI values were computed with the transition dipole moments from the converged CI expansion and extrapolated excitation energies. 
CSF-ROHF and CSF-ROKS results were computed using the nonorthogonal correction proposed in Ref.~\onlinecite{BourneWorster2021}. }
\label{tab:organic_radical_osc_strength}
\begin{ruledtabular}
\begin{tabular}{lcccccc}
& HF & LDA & PBE & PBE0 & $\omega$B97M-V & CIPSI \\
\hline
$[{+}{+}{-}]$ & 0.000 & 0.000 & 0.000 & 0.000 & 0.000 & 0.004 \\
$[{+}{-}{+}]$ & 0.390 & 0.622 & 0.635 & 0.647 & 0.676 & 0.139 \\
\end{tabular}
\end{ruledtabular}
\end{table}

The allyl radical is a more challenging case. 
CSF-ROKS excitation energies for seniority-1 Rydberg excitations are again in good agreement with literature theoretical best estimates (TBE), with $\omega$B97M-V achieving a MUE of \SI{0.03}{\eV}.
However, the valence $\mathrm{^2B_1}$ excited states are less well described across all methods. 
In contrast to vinyl, the near-degeneracy of the HOMO \textrightarrow\ SOMO and 
SOMO \textrightarrow\ LUMO excitations in allyl creates strong configurational mixing in the low-lying $\mathrm{^2B_1}$ 
valence states,\cite{Green2024} which cannot be recovered by a single-CSF approach.
To our knowledge, Ref.~\onlinecite{Green2024} provides the only literature values for the doublet 
HOMO \textrightarrow\ LUMO excitations, predicting both states to lie above \SI{9.5}{\eV} using generalized 
multiconfigurational quasi-degenerate perturbation theory. 
Other studies of allyl excited states do not report these states and the low-lying
$3^2A2$ state has been assigned to HOMO \textrightarrow\ $\mathrm{3p}$ character.\cite{Mach2010, Aquilante2003}
We also note that literature values for the excited $\mathrm{^2A_2}$ states are significantly spin contaminated,\cite{Mach2010} 
which may affect the accuracy of these reference values.
Regardless, the triplet-coupled HOMO \textrightarrow\ $\mathrm{3p}$ excitation energies are more accurate for all functionals 
compared to pure CSF-ROHF theory, demonstrating the accuracy gained by including an exchange-correlation functional.
A more accurate description of states with configurational mixing (e.g., the $\mathrm{^2B_1}$ states) may require 
a nonorthogonal CI expansion of multiple CSF-ROHF solutions or a future analogous expansion of CSF-ROKS states.

\subsection{Spin-flip excitation in molecular ruby}

\begin{figure}[h!]
\includegraphics[width=\linewidth,trim=0pt 0pt 5pt 0pt,clip]{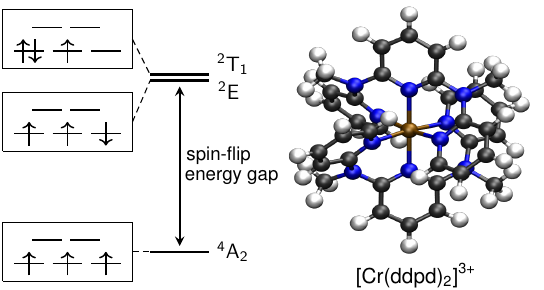}
\caption{The lowest excited states in the molecular ruby correspond to a spin-flip in the t$_2$ orbitals. 
In reality, the $\mathrm{^2E}$ and $\mathrm{^2T_1}$ microstates are interleaved in energy because the 
geometry of the complex is not strictly octahedral.}
	\label{fig:ruby}
\end{figure}

Next, we consider the excited states of the ``molecular ruby'' \ce{[Cr(ddpd)_2]^{3+}} complex 
(ddpd = $N,N'$-dimethyl-$N,N'$-dipyridin-2-ylpyridine-2,6-diamine), 
which has been shown to be a strongly near infrared luminescent complex.\cite{Otto2015,Heinze2026}
This complex is a d\textsuperscript{3} spin-flip emitter that undergoes phosphorescence from the
doublet $\mathrm{^2E}$/$\mathrm{^2T_1}$ excited states to the quartet $\mathrm{^4A_2}$ 
ground state (Fig.~\ref{fig:ruby}).
Crucially, this phosphorescence relies on the  $\mathrm{^2E}$/$\mathrm{^2T_1}$ states being 
the lowest-energy excited states to avoid  non-radiative internal conversion via the quartet manifold.

We consider the spin-flip vertical energy gap $\mathrm{^4A_{2}}\rightarrow \mathrm{^2E}$, which involves
transitions between states in which all the t\textsubscript{2} orbitals are singly occupied.
We expect that the quartet ground state $\mathrm{^4A_{2}}$  and the doublet $\mathrm{^2E}$ excited state
can be represented by a single CSF, using the $[{+}{+}{+}]$ and $[{+}{-}{+}]$/$[{+}{+}{-}]$ spin-coupling patterns respectively.
In contrast, the $\mathrm{^2T_1}$ configurations rely on a linear combination of configurations where one
of the t\textsubscript{2} orbitals is doubly occupied, which cannot be described by CSF-ROKS and is thus not considered here.
In reality, there can be some mixing between the $[{+}{-}{+}]$ and $[{+}{+}{-}]$ configurations that is beyond the 
description of CSF-ROKS, and thus we also consider the averaged energy of the $\mathrm{^2E}$ states, 
which is invariant to this mixing, as well as their splitting.
We used the high-spin optimized ROHF orbitals as an initial guess for
CSF-ROHF calculations with the $[{+}{-}{+}]$ and $[{+}{+}{-}]$ spin coupling, which were in turn used to 
initialize the corresponding CSF-ROKS calculations with a density functional approximation.
We employed the L-SR1 optimization approach to avoid any symmetry-broken local minima.

Table~\ref{tab:ruby} compares the CSF-ROKS excitation energies for a range of functionals against reference 
CASSCF and NEVPT2 results using the minimal (3e,5o) active space comprising the \ce{Cr} 3d orbitals,
and experimental values using the assignments reported in Ref.~\onlinecite{Reichenauer2024} and reviewed in Ref.~\onlinecite{Heinze2026}.
These data show that CSF-ROHF theory without any functional predicts excitation energies in close agreement with  CASSCF, 
indicating that these states can be accurately represented by a single CSF.
However, CSF-ROHF and CASSCF overestimate the energy gap by around \SI{0.9}{\eV} and \SI{0.8}{\eV}, respectively, 
as estimated using the $\mathrm{^2E}$ average.

\begin{table}[b!]
\caption{CSF-ROKS excitation energies for the $\mathrm{4A_2} \rightarrow \mathrm{^2E}$ transition in the \ce{[Cr(ddpd)_2]^{3+}}
complex using a range of functionals.
\textcolor{blue}{\textdagger}\ NEVPT2(7e,12o) energies are reported in Ref.~\onlinecite{Reichenauer2024} with a scaling of 0.89, 
which has been undone here to provide a direct theoretical comparison.}
\label{tab:ruby}
\begin{ruledtabular}
\begin{tabular}{lcccc}
           & \multicolumn{4}{c}{$\Delta E(\mathrm{^2E})\,/\,\text{eV}$} \\ \cline{2-5}
           & $[{+}{+}{-}]$ & $[{+}{-}{+}]$ & Average & Splitting \\
\hline
HF                           & $2.633$ & $2.634$ & $2.634$ & $0.001$ \\ 
LDA                        & $1.460$ & $1.541$ & $1.501$ & $0.081$\\
PBE                         & $1.248$ & $1.293$ & $1.270$ & $0.045$ \\
B97M-V                 & $1.823$ & $1.922$  & $1.872$ & $0.099$ \\
B3LYP                    & $1.481$ & $1.504$ & $1.492$ & $0.023$ \\
PBE0                       & $1.611$ & $1.634$	 & $1.622$ & $0.023$  \\
$\omega$B97M-V & $1.652$ & $1.681$ & $1.666$ & $0.029$\\
\hline
CASSCF(3e,5o)     & $2.502$ & $2.625$ & $2.564$ & $0.123$ \\
NEVPT2(3e,5o)     & $1.499$ & $1.618$ & $1.558$  & $0.119$ \\
NEVPT2(7e,12o)\cite{Reichenauer2024}   
& \hphantom{\textdagger}$1.934$\textsuperscript{\textcolor{blue}{\textdagger}}  
& \hphantom{\textdagger}$2.006$\textsuperscript{\textcolor{blue}{\textdagger}}  
& $1.970$ &  0.072 \\       
\hline  
Expt.\cite{Reichenauer2024}    & $1.686$ & $1.789$ & $1.738$ & $0.103$
\end{tabular}
\end{ruledtabular}
\end{table}

Introducing a density functional in CSF-ROKS consistently reduces the excitation energies by 
preferentially stabilizing the doublet state, bringing the predictions in closer agreement with the 
experimental results.
In particular, for the rung 4 hybrid (meta-)GGA functionals, the error in the $\mathrm{^2E}$ average ranges 
from \SI{-0.246}{\eV} for B3LYP to \SI{-0.072}{\eV} for $\omega$B97M-V, demonstrating 
that good agreement with the experimental results can be achieved for a suitable functional.
This reduction in the excitation energy after including dynamic correlation is consistent with previous 
NEVPT2 results using the (7e,12o) active space\cite{Reichenauer2024} 
and our own NEVPT2 results using the minimal (3e,5o) active space.
(Note that the NEVPT2 excitation energies reported in Ref.~\onlinecite{Reichenauer2024} were scaled by 0.89 
to give better agreement with experiment; we have undone that scaling to give a direct theoretical comparison.)
In all CSF-ROKS cases, the splitting of the $\mathrm{^2E}$ excitations is 
underestimated, although the B97M-V functional provides a remarkably good estimate with an error of \SI{-0.004}{\eV}, which is probably fortuitous.

These data demonstrate that CSF-ROKS can be used to predict different metal-centered
spin states in mononuclear transition metal complexes. 
This theory predicts the correct ordering of the quartet and doublet states with singly occupied 
t$_2$ orbitals in the molecular ruby, and provides a good estimate of the spin-flip energy gap.
The results are sensitive to the choice of functional, but this is common across DFT-based methods.
Regardless, we believe that CSF-ROKS is a promising spin-pure method 
for predicting low-spin open-shell states in transition metal complexes such as the molecular ruby.

\subsection{Bridged transition metal clusters}

Finally, we consider the stabilization of antiferromagnetic states in bridged transition metal clusters,
where low-spin ground states are known to be stabilized by kinetic exchange mechanisms involving the bridging ligands.\cite{Anderson1959,Malrieu2014}
These mechanisms are missing from the single CSF description due to the absence of any metal-to-ligand
charge transfer configurations, meaning that CSF-ROHF theory fails to predict low-spin 
ground states.\cite{Gouveia2024,Burton2025,Song2026}
However, it has been shown that optimizing the CSF-ROHF orbitals in the presence of a second-order
perturbation correction can correct the ordering of high- and low-spin states and achieve 
CASSCF-quality energetics.\cite{Song2026}
Here, we assess CSF-ROKS for a benchmark set of bridged transition-metal dimer clusters considered in Refs.~\onlinecite{Mayhall2014} and \onlinecite{Schurkus2020}. 
In each case, the individual metal atoms are prepared in a local high-spin configuration depending
on their d-orbital configuration, and these two local high-spin states are coupled either ferromagnetically 
to a high-spin state or antiferromagnetically to a low-spin state.

First, we consider the iron-sulfur dimer \ce{[Fe2S2(SCH3)_4]^{2-}}, which has previously been 
studied using CSF-ROHF theory.\cite{Gouveia2024,Burton2025,Song2026}
For every exchange-correlation functional considered, we have been able to converge an
antiferromagnetic solution representing the low-spin state.
However, in all cases, we find that CSF-ROKS fails to correctly predict a low-spin ground state,
 as shown in Table~\ref{tab:FeS}.
Furthermore, the energy gap $\Delta E_{\text{LS-HS}} = E_{\text{LS}} - E_{\text{HS}}$ for all the 
DFT functionals is substantially more positive than the HF description itself, although it is closer
to the HF result when some exact HF exchange is included in hybrid functionals.

\begin{table}[htb!]
\caption{High-spin $[{+}{+}{+}{+}{+}{+}{+}{+}{+}{+}]$ and low-spin $[{+}{+}{+}{+}{+}{-}{-}{-}{-}{-}]$ CSF-ROKS energies for the \ce{[Fe2S2(SCH3)4]^2-} cluster with the cc-pVDZ basis set. 
All the exchange-correlation functionals that were tested incorrectly predict a high-spin ground state.}
\label{tab:FeS}
\begin{ruledtabular}
\begin{tabular}{lccc}
 & $E_{\text{LS}} / \mathrm{E_h}$ & $E_{\text{HS}} / \mathrm{E_h}$ & $\Delta E_{\text{LS-HS}} / \mathrm{cm^{-1}}$ \\
\hline
HF              & $-5068.7551$ & $-5068.7592$ & $910$  \\ 
LDA             & $-5047.7974$ & $-5047.8129$ & $3401$ \\
PBE             & $-5074.8281$ & $-5074.8407$ & $2771$ \\
PBE0            & $-5075.0603$ & $-5075.0718$ & $2533$ \\
$\omega$B97M-V  & $-5076.8741$ & $-5076.8851$ & $2419$
\end{tabular}
\end{ruledtabular}
\end{table}

\begin{table*}[htb!]
	\caption{Exchange coupling constants computed using the Yamaguchi formula [Eq.~\eqref{eq:Jcoupling}] for CSF-ROKS using various  functionals, compared to broken-symmetry CCSD(T) and UKS data using the cc-pVDZ basis set. 
\textcolor{blue}{\textdagger}\ UKS, CCSD(T), and experimental data are taken from Ref.~\onlinecite{Schurkus2020} and references therein.}
\label{tab:Jcoupling}
	\begin{ruledtabular}
		\begin{tabular}{lcccccccc}
			\multirow{2}{*}{$J\, /\, \mathrm{cm}^{-1}$} &  \multicolumn{5}{c}{CSF-ROKS } &   & & 
			\\ \cline{2-6}
			& HF & LDA & PBE & PBE0 & $\omega$B97M-V  & UKS (PBE)\textsuperscript{\textcolor{blue}{\textdagger}}  & CCSD(T)\textsuperscript{\textcolor{blue}{\textdagger}}   &  Expt.\textsuperscript{\textcolor{blue}{\textdagger}}  \\
			\hline
			\ce{[Cr2O(NH3)10]^4+}   & $15$  & $59$  & $49$   & $39$ & $34$ & $-356$ & $-194$  & $-225$ \\ 
			\ce{[Mn2O(CN)10]^6-}     & $23$ & $124$ & $100$ & $78$ & $65$ & $-2085$ & $-167$ & $-360$ \\ 
			\ce{[Fe2OCl6]^2-}            & $28$ & $91$   & $77$   & $75$ & $66$ & $-246$ & $-123$   & $-112$ \\ 
			\ce{[Fe2S2(SCH3)4]^2-} & $30$ & $113$  & $92$  & $84$ & $81$ & $-850$ & $-191$   & $-148$ \\ 
			\ce{[Cu2Cl6]^2-}              & $12$ & $185$  & $123$ & $82$ & $63$ & $-140$ & $-12$  & $-19$ \\ 
		\end{tabular}
	\end{ruledtabular}
\end{table*}

This failure is repeated across all the transition metal dimers considered here.
Since the benchmark systems have different numbers of unpaired metal-centered d-electrons, 
we report the $J$ coupling constant using the Yamaguchi formula\cite{Yamaguchi1986}
\begin{equation}
J=\frac{E[\Psi_\text{HS}] - E[\Psi_\text{LS}]}{ \expval*{\hat{S}^2}_{\Psi_\text{LS}} - \expval*{\hat{S}^2}_{\Psi_\text{HS}} },
\label{eq:Jcoupling}
\end{equation}
where we compute the high-spin and low-spin energies using spin-pure CSF-ROKS theory.
In all cases, CSF-ROKS erroneously predicts a positive $J$ coupling constant, 
indicating a high-spin ferromagnetic ground state (Table~\ref{tab:Jcoupling}). 
In contrast, previous results from UKS theory and broken-symmetry coupled cluster correctly
predict a low-spin antiferromagnetic ground state
in agreement with experiment.\cite{Schurkus2020}

These results indicate that the current formulation of CSF-ROKS cannot capture the 
kinetic exchange mechanisms required to correctly predict antiferromagnetic ground states
in bridged transition metal complexes.
We believe that these effects might be captured by including a perturbation
correction using a double-hybrid functional, but we were unable to test this approach due to the 
absence of such functional implementations in PySCF.

\section{Conclusions}

We have introduced the CSF-ROKS method as a generalization of ROKS theory to 
an arbitrary genealogical spin-coupling pattern.
The key to our approach is to find a suitable ensemble of unrestricted Slater determinants
that exactly recovers the Hartree--Fock exchange energy of the corresponding spin-coupled CSF.
We have shown that the number of determinants required to construct this ensemble scales
only quadratically with the number of open shells in the genealogical CSF.
The restricted ensemble orbitals can then be optimized using our recently developed 
geometric direct minimization approach for CSF-ROHF optimization,\cite{Burton2025} providing a 
generalized spin-pure approach to low-spin open-shell KS-DFT.

Our numerical results for the \ce{N2} binding curve show that CSF-ROKS provides a meaningful
extension of CSF-ROHF theory by introducing dynamic correlation 
without losing the chemical character of individual spin-coupled configurations.
Crucially, the fully bond-broken spin-coupling pattern $[{+}{+}{+}{-}{-}{-}]$ dissociates 
to the correct $2\times\text{N}$ atomic limit, demonstrating size consistency.
Using the diabatic character of different spin-coupled configurations, we have shown how
DFT-based dynamic correlation preferentially stabilizes low-seniority configurations in the 
polyacene series, delaying the onset of polyradical character to longer acene chains compared
to standard CSF-ROHF theory.
Furthermore, CSF-ROKS can be used to target spin-pure state-specific
excited states, as demonstrated for up to three unpaired electrons in monoradicals and
metal-centered spin-flip excitations in the \ce{[Cr(ddpd)_2]^{3+}} molecular ruby complex.
However, our current formulation cannot capture the kinetic exchange mechanisms 
that stabilize low-spin antiferromagnetic ground states in bridged transition metal clusters.

The principal advantage of the CSF-ROKS approach is that it can target spin-pure
open-shell states with any spin-coupling pattern, avoiding issues with interpreting 
spin-polarized UKS results that may be spin-contaminated.
Moving forwards, we anticipate that CSF-ROKS may provide a useful 
foundation to develop linear response theories for dynamically-correlated neutral 
and charged excitations in low-spin open-shell molecules, in an analogous way to
recent developments for CSF-ROHF theory.\cite{Gouveia2025}
Furthermore, the low-order computational scaling of CSF-ROKS compared to correlated 
multiconfigurational methods may enable accurate geometry optimization for
low-spin open-shell molecules or provide spin-pure data to train
machine-learned interatomic potentials.

In its current formulation, CSF-ROKS is unable to predict low-spin open-shell
ground states that are stabilized by  kinetic exchange\cite{Anderson1959} or 
spin-polarization\cite{Kollmar1978} mechanisms.
However, we expect that it may be possible to recover some of these effects using a
second-order perturbative correction to the CSF-ROKS description.
This correction could be applied to the CSF wavefunction in 
an analogous manner to the split-GAS approach of Ref.~\onlinecite{Song2026}, 
or it could be introduced into the CSF-ROKS ensemble using a double hybrid functional.
Whether these improvements to CSF-ROKS are sufficient to predict antiferromagnetic 
ground states is left for future research.

\appendix

\section{Gradient of the CSF-ROKS energy}
\label{sec:gradient}
In this section, we derive the gradient for the CSF-ROKS energy provided in Eq.~\eqref{eq:CSFROKSgrad}.
Since the terms arising from the one-electron terms and Coulomb interaction are the same as standard 
CSF-ROHF theory (see Ref.~\citenum{Burton2025}), we focus only on contributions from the exchange-correlation 
energy Eq.~\eqref{eq:zieglersumXC} and assume real-valued orbitals.
The task is therefore to evaluate the derivatives $\frac{\delta E_\xc}{\delta \kappa_{pq}}$.
First, from the linearity of the ensemble energy, we obtain
\begin{equation}
\frac{\delta E_\xc}{\delta \kappa_{pq}} = \sum_{L} c_L \frac{\delta E_\xc[\Phi_L]}{\delta \kappa_{pq}}.
\label{eq:xcgrad}
\end{equation}
We follow Refs.~\citenum{Gould2021} and \citenum{Lehtola2020} by assuming that the exchange-correlation
energy is a functional of the one-body density matrices for each spin
\begin{equation}
\gamma_\sigma^L(\br,\br') = \sum_m n^L_{m\sigma} \phi_m(\br) \phi_m(\br'),
\end{equation}
which is true for all commonly used functionals.
These density matrices can also be expressed in the AO basis as
\begin{equation}
\gamma_\sigma^L(\br,\br') = \sum_{\mu \nu} P^{L}_{\mu \nu,\sigma} \chi_\mu(\br) \chi_\nu(\br'),
\end{equation}
where the density matrix elements for determinant $\Phi_L$ are
\begin{equation}
P^{L}_{\mu \nu,\sigma} = \sum_{m} n^L_{m\sigma} C_{\mu m} C_{\nu m}.
\end{equation}
The individual contributions from each determinant in the ensemble can be evaluated
using the chain rule as
\begin{equation}
\frac{\delta E_\xc[\Phi_L]}{\delta \kappa_{pq}} 
= \sum_{\sigma \mu \nu} \frac{\delta E_\xc[\Phi_L]}{\delta P^{L}_{\mu \nu,\sigma}}  \frac{\delta  P^{L}_{\mu \nu,\sigma}}{\delta \kappa_{pq}} .
\label{eq:detGrad}
\end{equation}
The matrix elements of the exchange-correlation potential in the AO basis are defined as\cite{Lehtola2020} 
\begin{equation}
\mel{\chi_\mu}{\hat{v}_{\xc,\sigma}^L}{\chi_\nu} =  \frac{\delta E_\xc[\Phi_L]}{\delta P^{L}_{\mu \nu,\sigma}}
\label{eq:AOvxc}
\end{equation}
and the derivatives of the density matrix elements are
\begin{equation}
\begin{split}
\frac{\delta P_{\mu \nu,\sigma}^L}{\delta \kappa_{pq}} &= 2\sum_m n_{m \sigma}^L \frac{\delta C_{\mu m}}{\delta \kappa_{pq}} C_{\nu m}
\\
&= 2\qty(n_{q\sigma}^L C_{\mu p} C_{\nu q} - n_{p\sigma}^L C_{\mu q}C_{\nu p}).
\end{split}
\label{eq:densDeriv}
\end{equation}
Here, we have exploited the symmetry of the density matrix for real orbitals and the derivative of the orbital coefficients
under the exponential transformation [Eq.~\eqref{eq:expK}] as
\begin{equation}
\frac{\delta C_{\mu m}}{\delta \kappa_{pq}} = \delta_{qm}C_{\mu p} - \delta_{pm} C_{\mu q}.
\label{eq:orbDeriv}
\end{equation}
%
%
Inserting Eqs.~(\ref{eq:AOvxc}--\ref{eq:orbDeriv}) into Eq.~\eqref{eq:detGrad} then yields
\begin{equation}
\frac{\delta E_\xc[\Phi_L]}{\delta \kappa_{pq}} 
= 
\sum_{\sigma} 2\qty(n_{q\sigma}^L \mel{\phi_q}{\hat{v}^L_{\xc,\sigma}}{\phi_p} - n_{p\sigma}^L \mel{\phi_p}{\hat{v}^L_{\xc,\sigma}}{\phi_q}),
\label{eq:Lgrad}
\end{equation}
where we have applied the hermitian symmetry $\mel{\phi_q}{\hat{v}^L_{\xc,\sigma}}{\phi_p}=\mel{\phi_p}{\hat{v}^L_{\xc,\sigma}}{\phi_q}$
for notational convenience.
Combining Eq.~\eqref{eq:Lgrad} and Eq.~\eqref{eq:xcgrad} then gives
\begin{align}
\frac{\delta E_\xc}{\delta \kappa_{pq}} 
&= 
\sum_{L\sigma} 2 c_L \qty(n_{q\sigma}^L \mel{\phi_q}{\hat{v}^L_{\xc,\sigma}}{\phi_p} - n_{p\sigma}^L \mel{\phi_p}{\hat{v}^L_{\xc,\sigma}}{\phi_q}),
\nonumber
\\
&= 2\qty(\mel{\phi_q}{\hat{v}^\cQ_{\xc}}{\phi_p} - \mel{\phi_p}{\hat{v}^\cP_{\xc}}{\phi_q}).
\label{eq:gradxc_final}
\end{align}
where we have defined
\begin{equation}
\mel{\phi_m}{\hat{v}^\cQ_{\xc}}{\phi_n} =  \sum_{L\sigma} c_L n_{q\sigma}^L \mel{\phi_m}{\hat{v}^L_{\xc,\sigma}}{\phi_n}.
\label{eq:effVxc}
\end{equation}
by noting that, for every determinant in the ensemble,
the spin-orbital occupation numbers $n_{q\sigma}^L$ are the same for all orbitals in the same shell $q \in \cQ$.
The final expression Eq.~\eqref{eq:gradxc_final} provides the exchange-correlation contribution to the 
CSF-ROKS energy gradient in Eq.~\eqref{eq:CSFROKSgrad} via its inclusion in the generalized Fock matrix elements 
Eq.~\eqref{eq:gFock}.

\section{Hessian of the CSF-ROKS energy}
\label{sec:hessian}
We now derive the approximate orbital Hessian of the CSF-ROKS energy used to construct the 
preconditioner in Eq.~\eqref{eq:prec}.
The diagonal components of the orbital Hessian $Q_{pq,pq}$ are defined in Eq.~\eqref{eq:hess}.

The one-electron and Coulomb energy components give rise to contributions through the $n_\cP \qty(h_{pq} + J_{pq})$
term in the generalized Fock matrix [Eq.~\eqref{eq:gFock}] and the term $4(a_{pp}+ a_{qq} - 2a_{pq}) \braket{pq}{qp}$ 
in Eq.~\eqref{eq:prec}.
Any (scaled) exact HF exchange also contributes through the $\xi K^{\cP}_{pq}$ term in the generalized 
Fock matrix and the term $2\xi(b_{pp} + b_{qq} - 2b_{pq}) (\braket{pq}{qp} + \braket{pq}{pq})$ in Eq.~\eqref{eq:prec}.
The derivation of these terms is the same as for CSF-ROHF in Ref.~\citenum{Burton2025}, 
and thus no further detail is provided here.

The exchange-correlation contribution can be derived through the linearity of the ensemble energy as
\begin{equation}
\frac{\delta^2 E_\xc}{\delta \kappa_{pq} \delta \kappa_{pq}} = \sum_L c_L \frac{\delta^2 E_\xc[\Phi_L]}{\delta \kappa_{pq} \delta \kappa_{pq}} .
\label{eq:deriv2ensemble}
\end{equation}
The contribution from each determinant is  given by the derivative of Eq.~\eqref{eq:detGrad} as
\begin{equation}
\begin{split}
\frac{\delta^2 E_\xc[\Phi_L]}{\delta \kappa_{pq} \delta \kappa_{pq}} =
& \sum_{\substack{\sigma \tau \\\mu \nu \lambda \gamma}} 
\frac{\delta^2 E_\xc[\Phi_L]}{\delta P^{L}_{\lambda \gamma,\tau} \delta P^{L}_{\mu \nu,\sigma}}   
\frac{\delta  P^{L}_{\lambda \gamma,\tau}}{\delta \kappa_{pq}}  
\frac{\delta  P^{L}_{\mu \nu,\sigma}}{\delta \kappa_{pq}} 
\\
&+ \sum_{\mu \nu \sigma} 
\frac{\delta E_\xc[\Phi_L]}{\delta P^{L}_{\mu \nu,\sigma}}  \frac{\delta^2  P^{L}_{\mu \nu,\sigma}}{\delta \kappa_{pq}\delta \kappa_{pq}} .
\end{split}
\end{equation}
Since we only require an approximation to the diagonal Hessian matrix elements to define the preconditioner, we ignore 
the second functional derivatives of the exchange-correlation energy such that
\begin{equation}
\frac{\delta^2 E_\xc[\Phi_L]}{\delta \kappa_{pq} \delta \kappa_{pq}} \approx \sum_{\mu \nu \sigma} 
\frac{\delta E_\xc[\Phi_L]}{\delta P^{L}_{\mu \nu,\sigma}}  \frac{\delta^2  P^{L}_{\mu \nu,\sigma}}{\delta \kappa_{pq}\delta \kappa_{pq}}.
\label{eq:detGrad2}
\end{equation}
The second derivatives of the spin-density matrices are 
\begin{align}
\frac{\delta^2 P_{\mu \nu,\sigma}^L}{\delta \kappa_{pq}\delta \kappa_{pq}} &= 
2\sum_m n_{m \sigma}^L \qty(\frac{\delta^2 C_{\mu m}}{\delta \kappa_{pq}\delta \kappa_{pq}} C_{\nu m} +
\frac{\delta C_{\mu m}}{\delta \kappa_{pq}} \frac{\delta C_{\nu m}}{\delta \kappa_{pq}})
\nonumber \\
&= 2\qty(n_{p\sigma}^L - n_{q\sigma}^L)\qty(C_{\mu q} C_{\nu q} - C_{\mu p}C_{\nu p}),
\label{eq:densDeriv2}
\end{align}
where we have used the second derivatives of the orbital coefficients under the exponential transformation [Eq.~\eqref{eq:expK}] 
\begin{equation}
\frac{\delta^2 C_{\mu m}}{\delta \kappa_{pq}\delta \kappa_{pq}} = - \delta_{pm} C_{\mu p}  - \delta_{qm}C_{\mu q}.
\end{equation}
Inserting Eqs.~\eqref{eq:densDeriv2} and \eqref{eq:AOvxc} into Eq.~\eqref{eq:detGrad2} then yields
\begin{equation}
\begin{split}
\frac{\delta^2 E_\xc[\Phi_L]}{\delta \kappa_{pq} \delta \kappa_{pq}} &\approx 
\\
&\sum_\sigma
2(n_{p\sigma}^L - n_{q\sigma}^L)\qty( \mel{\phi_q}{\hat{v}^L_{\xc,\sigma}}{\phi_q}) - \mel{\phi_p}{\hat{v}^L_{\xc,\sigma}}{\phi_p}),
\end{split}
\end{equation}
and evaluating the summation over all the ensemble determinants using Eq.~\eqref{eq:effVxc} leads to 
\begin{equation}
\begin{split}
\frac{\delta^2 E_\xc}{\delta \kappa_{pq} \delta \kappa_{pq}} \approx\
&2\qty(\mel{\phi_q}{\hat{v}^\cP_{\xc}}{\phi_q} -\mel{\phi_q}{\hat{v}^\cQ_{\xc}}{\phi_q}) 
\\ 
+\ &2\qty(\mel{\phi_p}{\hat{v}^\cQ_{\xc}}{\phi_p} -\mel{\phi_p}{\hat{v}^\cP_{\xc}}{\phi_p}).
\end{split}
\end{equation}
Finally, we note that these terms contribute to the approximate diagonal Hessian matrix elements 
in Eq.~\eqref{eq:prec} through the definition of the generalized Fock matrix elements defined in Eq.~\eqref{eq:gFock}.

\section*{Data availability}
All numerical data produced in this study and the molecular geometries used are available in a public repository at \href{https://doi.org/10.5281/zenodo.24438513}{10.5281/zenodo.24438513}.

\section*{Conflicts of Interest}
The authors declare no conflicts of interest.

\section*{Acknowledgements}
H.G.A.B.\ is a Royal Society University Research Fellow (URF\textbackslash R1\textbackslash 241299) at University College London.

\section*{References}
\bibliography{manuscript}
\end{document}